\documentclass{article}
\usepackage{pgfplots}
\usepackage[margin=1in]{geometry}
\usepackage{natbib}
\usepackage{multicol}

\title{Conversational User Interfaces for Blind Knowledge Workers \\
\large{A Case Study}}
\author{
    Kyle Dent\\
    kdent@parc.com\\
    Palo Alto Research Center
  \and
    Kalai Ramea\\
    kramea@parc.com\\
    Palo Alto Research Center
}

\begin{document}

\maketitle

\begin{abstract}
Modern trends in interface design for office equipment using controls on touch surfaces create greater obstacles for blind and visually impaired users and contribute to an environment of dependency in work settings. We believe that \textit{conversational user interfaces} (CUIs) offer a reasonable alternative to touchscreen interactions enabling more access and most importantly greater independence for blind knowledge workers. We present a case study of our work to develop a conversational user interface for accessibility for multifunction printers. We also describe our approach to conversational interfaces in general, which emphasizes task-based collaborative interactions between people and intelligent agents, and we detail the specifics of the solution we created for multifunction printers. To guide our design, we worked with a group of blind and visually impaired individuals starting with focus group sessions to ascertain the challenges our target users face in their professional lives. We followed our technology development with a user study to assess the solution and direct our future efforts. We present our findings and conclusions from the study.
\end{abstract}

\section{Introduction}

Manufacturers of multifunction printers (MFP) proudly tout their latest touchscreen user interfaces that minimize the physical controls while allowing more complex customization. These new UIs let users configure workflows that are available with a few taps or swipes on built-in surfaces much like personal tablet devices.  The machines look sleeker and permit more complicated interactions than were easily available with mechanical push buttons. Unfortunately, blind and visually impaired office workers can no longer use machines with surface-only interfaces. In the past, blind workers could learn to feel for the physical buttons and accomplish many of the tasks required in their jobs. Learning a single model of machine, its functions, and where its buttons were located was never an ideal option, but it at least allowed some degree of independence. With the newest interfaces on office equipment, blind users have, for the most part, been left out. To address this issue, our research group
at the Palo Alto Research Center (formerly Xerox PARC)
developed a conversational user interface that provides access to most of the important features that simply wouldn't otherwise be available to these users without assistance.

Our solution was built for the latest family of Xerox MFPs that expose an API for programmatic control over their operations. The machines we worked with offer five high-level functions: print, scan, copy, fax, and email. Our conversational agent does not support printing since that function is mostly driven from a computer or other device that already offers assistive features. The targeted use-case is for blind and visually impaired individuals standing at a copier who want to make a copy, scan a document or send a fax or email. Many machine options such as single- or double-sided copying, number of copies, binding choices, and image variations like lightening or darkening are available. The combination of choices and options across all the functions amounts to nearly a hundred possible selections. Our solution covers about half of these and prioritizes the most frequently used options. While there might have been good alternative multi-modal interface designs that make use of the screen to, for example, display large icons with high contrast to help the visually impaired, our goal was to create a solution that serves people with total blindness, requiring all functions to be accessible conversationally.

\section{Challenges in Spoken Interfaces}
\label{sec-challenges}

Conversational and voice interactions have become more common recently, and research has followed the new technology to understand how people react and interact with it. In one study, [\citet{Porcheron:2018:VIE:3173574.3174214}] collected data from family interactions with standalone, screenless smart speakers in the home. Their paper points out that while these devices are marketed to help customers get things done, little is known about what people are actually able to accomplish with their devices. From an ethnomethodological perspective, using Conversation Analysis, the researchers are trying to understand how conversational devices fit into human interactions and what is being achieved within these dialogs. They were able identify very little actual collaboration between people and their devices. The collaboration they did find related to the mechanics of the conversation rather than in service of accomplishing an external goal.

Indeed, we have noticed that task-based, collaborative, conversational interfaces, as opposed to question answering or transaction processing, continue to be rare. In contrast, our research interest is in developing interactions between people and technology operating together on a shared goal. In addition, we want to distinguish our interface from one that simply maps voice commands to individual functions on a machine. Having the ability to carry on rich and flexible conversations allows for more complicated tasks and composed workflows.

Developing effective conversational user interfaces requires special skills and can be difficult to implement successfully. Moreover, the previously mentioned trend to put graphical user interfaces on office equipment provides sighted users a very familiar interaction paradigm. But GUIs differ in many ways from conversational interfaces, most significantly in how they model tasks for users. By design GUIs dictate the way that tasks are done, but people are quick to adapt to them. Conversational user interfaces (CUIs), on the other hand, must adapt to users with differing mental models of the task to be performed. Because users will sometimes take the initiative in the interaction, CUIs must  adapt responses to fit the current context. They must also accept many different phrasings of the same thing in addition to handling inconsistencies of language and general disfluencies.

Managing the mechanics of conversation is actually a secondary consideration
for task-based conversational agents. Intelligent assistants must also determine the goal of the
task and move the dialog towards accomplishing it. For our solution, we have
drawn from the work presented in [\citet{Grosz:1986:AIS:12457.12458}] for our
understanding of dialog, especially with regard to task-based, collaborative
interactions such as the one described in the expert/apprentice flywheel example first discussed in
[\citet{Grosz:1978:FD:980262.980278}]. We have also depended on other more recent work
that considers dialog as an interface modality both for accessibility and
hands-free/eyes-free use cases. Across all of the research on CUIs, it is
apparent that there continue to be many challenges.

\subsection{Problems specific to CUIs}
[\citet{Corbett:2016:ISA:2935334.2935386}] describes the problem of discoverability, constraint, and affordances in their voice user interface for controlling Android devices. Their interest was an interface for individuals with limited hand dexterity, meaning visual clues and presentation were still possible to alleviate some of the problems inherent in CUIs. [\citet{Yankelovich:1996:UKS:242485.242500}] also described the problem of discoverability and pointed out that speech interfaces are like command-line user interfaces in that functions are hidden. GUIs were invented in large part to bring hidden functionality to the surface. In our case, for example, many users are not aware that MFPs may have a built-in stapler or other binding options. Even the existence of an option to choose double-sided versus single-sided printing may not be apparent to a new user.

Yankelovich goes on to point out the additional problem in conversational interfaces where users who, once they start speaking, assume the system has capabilities that it  does not. Her approach is to provide carefully formed voice prompts to limit how much free expression is available to users. We have tried to adopt this principle but only to the extent that it won't conflict with our goal of keeping the system conversational. For example, we make use of her suggested implicit confirmations where possible, so that the interaction can flow more naturally without conversational turns dedicated specifically to confirming information.

\subsection{CUIs for blind users}
While there has been work that shows that personal assistants have been helpful
in providing access to technology that might otherwise be inaccessible to users
with various disabilities, there is also research indicating particular
problems in these devices for blind users. Among other issues,
[\citet{Abdolrahmani:2018:STY:3234695.3236344}] identified the problem of
appropriate feedback from voice systems. Many conversational devices rely on
visual cues to report certain events or states. For example, a light of some
kind is often used to indicate that a microphone is listening. Amazon's Echo
device relies on a light ring to indicate different kinds of activity on the
device. Similarly, small buttons for muting or initiating a conversation are
not easily found without vision. The authors also describe the lack of control
over voice outputs. Some individuals may require a slower delivery while many
others who are accustomed to screen readers set to very fast rates may find
listening to the responses frustrating.

[\citet{Gotzelmann:2017:PCA:3056540.3064954}] discusses issues of accessibility with 3D printers. Their work covers the challenges of physically interacting with equipment without vision, which aligns closely with our own problem of using an MFP. They break down the workflow for 3D printing into discrete steps. We follow the same approach and similarly face the problem they describe of a large number of parameters and alternatives that are available for each step of the process. They resolved the problem by reducing the number of options to those that are most essential in order to shrink the scope and complexity of the interaction. Our ambition, on the other hand, is to provide access for blind users to all of the options that are available on MFPs. Unfortunately, there were technical limitations in how our agent communicates with the device making our ideal impossible using our current architecture.

\section{Focus Groups and Usability}

To understand more about the experiences of our target users, we engaged with
two different user groups. The first group we met with was the Association for the
Blind and Visually Impaired (ABVI) in Rochester, New York. ABVI provides
services to people with significant vision loss including training for skills
needed for working and living independent lives.  Our informal visit with them
helped us to gain an initial understanding of the current situation for office
workers with visual impairment, seeing some of the assistive devices available,
and learning about challenges they face day-to-day in their work.

Subsequently, we partnered with the Vista Center for the Blind and Visually
Impaired in Palo Alto, California. The Vista Center assembled a user group
of eleven people who were current or past office workers with significant
experience using multifunction printers. Their ages ranged from 47 to 73 years
old with a median of 62.5. All of them live with moderate to severe visual
impairment with most having no functional sight.

We conducted a focus group to learn more about specific difficulties. Users
expressed, in addition to the problems of touchscreen interfaces, a lack of
accessible technology in their work environments resulting in reduced
independence and often exclusion from work teams. They complained about complex
choices across many options with poor navigational directions causing
confusion, inefficiency, and frustration.

Following are some examples from comments we collected during the focus group session.

\begin{quote}
\textit{
\ldots if you want it collated, double-sided, stapled, whatever and you
have to push the right buttons and all of those things, or otherwise you have
to have a sighted aide or another oral aide, or even a student aide{\ldots}and
a lot of the districts just don't have funding for a lot of that. Really
frustrating.}
\end{quote}

When errors happen, there is little or no explanation of the error or the
cause.

\begin{quote}
\textit{
I can't just look at the page to make sure it's printing properly. I'm relying
on just, knowing it is, and sometimes the first inkling that I know something
was wrong, is that I hand the page to somebody, and they say, ``Um, this only
has like two-thirds of the printing on it.'' Or, ``This is really light in
color.'' And then I know that our toner cartridge wasn't working right or two
pages got stuck together when the paper was feeding through, and it didn't feed
right}
\end{quote}

They also mentioned the difficulty of inconsistent designs across different
brands of devices all of which lack verbal introductions or tutorials.

\begin{quote}
\textit{
I have to ask somebody to make sure that the page is turned, so I know which
way\ldots}
\end{quote}

We heard about workarounds they employ to carry out many tasks. For example,
distinguishing which job is the right one in the output tray is a frequent
problem.

\begin{quote}
\textit{
\ldots [I] always check the output tray before I print anything, and I make
sure that it is empty.}
\end{quote}

\begin{table*}[ht]
\caption{Focus Group Interface Design Recommendations}
\begin{tabular}{p{3.0in} p{3.0in}}
\hline
Feature (``What'') & Reduces OR Solves (``Why'') \\
\hline
Invitation to set defaults &
Repetitive ``starting over'' for each job; UI disorientation; cognitive
overload from complex workflows; inefficiency \\

Invitation to choose a basic function, then add options &
Reduce cognitive overload from complex feature lists; frustration \\

``Auditory'' buttons and lexicon of sound cues &
Undesired output from mistaken input \\

Voice-over walkthrough for frequently used options &
Eliminates need to listen through lengthy menus \\

Invitation to ask for help at any time &
Avoids user ``bail-outs'' and job abandonment \\

Invitation for a new user to take a tour of the device and interface &
Facilitate new user orientation \\

User-prompted description of features, options, input process &
Reduce need for workarounds, enable use of full feature set, support correct
paper loading \\

Verbalization of error message/code, its meaning and troubleshooting support &
Less frustration and confusion; minimize device damage from inappropriate
troubleshooting \\

Verbal preview of output &
Avoid need for ``another pair of eyes'' enables independence \\

Query and confirmation for ``unusual'' or non-routine request &
Less frustration from undesired device outputs, minimize resource waste (e.g.,
unnecessary copies etc.) \\

Verbal update of job status/progress and output specification &
Avoid restarting job, resource waste, output collection error \\
\hline
\end{tabular}
\label{tbl-design-recommend}
\end{table*}

From the challenges and suggestions we heard, we developed a set of design
recommendations for our conversational agent shown in Table
\ref{tbl-design-recommend}.

\section{Design Goals and Principles}

Based on the lessons we learned from the focus group sessions and our own prior work on conversational agents
[\citet{Dent2018}],
we formulated several goals and principles to apply in the solution.

Creating a good conversational interaction that allows individuals to be
successful in task-based goals was a high priority. We wanted our agent to be
collaborative. Keeping task objectives at the center of the design, we wanted
the interactions to feel as normal as possible. We are aware of the risks of
trying to copy human-like behaviors in agents
[\citet{Chefitz:2018:DCI:3197391.3205439}], which we avoid. We are interested,
however, in modeling aspects of human-to-human conversation to improve our
interactions but without setting up unmeetable expectations. For example, human
conversation is strongly characterized by mixed-initiative where individuals
can seamlessly introduce new information or take the lead during a
conversation. Our agent allows for similar flexibility. Users can provide
information independent of system prompts to provide it. The agent must be
prepared to recognize information that arrives outside of a pre-planned
sequence.

We defined the following principles in developing our conversational design.

\begin{description}
\item[Be Accommodating] As explained, users should be able to provide
information whenever they want, and they should be able to take the initiative
in the conversation. They should be able to answer questions naturally with
either fragments or complete sentences.  They should be able to introduce and
ask for multiple things within a single turn. The agent should allow for lulls
in the conversation. When the agent has asked the user to perform an action, it
might take some time before the action is complete. The agent should wait until
users resume the conversation.

\item[Be Brief] Long utterances can be hard to follow and retain. When
listing options, for example, provide them in manageable chunks allowing users
to ask for more as they are ready.

\item[Be Helpful] If users have not provided required information to complete a
task, try to give other relevant information or suggest alternatives that
haven’t been mentioned before. If people ask for help, include the option to
give more information about related topics. Provide procedural help for new
users. When the agent speaks, expect conversational mirroring. The agent should
use language and structures that it can understand to give users clues about
the best way to say things to maximize the chances of recognition of future
utterances.

\item[Be Transparent] As much as possible convey to users what the agent is thinking and doing. If users' utterances are ambiguous or unclear, confirm the most likely interpretation. Give cues about where the collaboration is in a process (e.g. ``First\ldots'', ``Then\ldots'', ``Finally\ldots''). Use acknowledgments whenever users supply new information (e.g.  ``OK'', ``Got it''). Perform a final confirmation before invoking the printer, especially for jobs that might use a lot of resources (e.g. ``I need 500 copies of this document'').
\end{description}

\section{The Solution}

The solution consists of a device that includes a controller, a microphone, and a speaker. We used a Raspberry Pi for the controller and after trying several different microphones, settled on one designed for conference room phones. In the range of relatively inexpensive microphones, this option performed best to capture voice signals from anywhere around the printer.

\begin{figure*}
\includegraphics[scale=0.6]{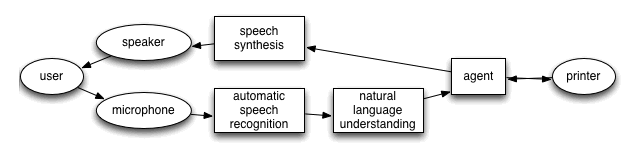}
\caption{An integrated conversational system to augment and support MFP functions comprised of custom-built hardware and software to
manage both conversational dialog strategy and also interaction with the device to operate its functions.}
\label{fig-arch}
\end{figure*}

\subsection{Architecture}
Spoken language is captured at the device and sent to Google Cloud
Speech-to-Text for automatic speech recognition processing. From Google's
service we obtain the textual representation which is fed into our our dialog
system for natural language understanding and processing by the dialog manager.
After making a decision about what to do next, the dialog manager communicates
with the multifunction device and sends its generated response to Amazon Polly
for speech synthesis. We also have a version that uses local software libraries for
speech recognition and synthesis that runs without depending on any cloud
services. Figure \ref{fig-arch} displays a schematic view of the architecture.

Before building our own hardware device, we experimented with commercially
available smart speakers.  We found that for interactions lasting more than a
couple of turns, interacting with these devices is awkward. The services
supplied by the manufacturers limit the amount of control
throughout the interaction. The devices also require a wake word, which we did
not prefer, and provide buttons that are too difficult to find, given limited
or no vision.

On our custom-made device, conversations with the system are initiated by
pressing a large button. We explored using a wake word on our own device as
well, but preferred the button to eliminate issues with wake word recognition.
Copiers tend to be in rather noisy environments, which makes the wake word
option even less reliable. Once the conversation is initiated, we open a
listening channel that stays open until the task is completed. This eliminates
the need to press the button on every turn allowing more natural conversational
interactions.

The open channel introduces other complications, however. The voice signal
processor has to determine when a user's turn ends, and side conversations that
are not directed at the agent have to be distinguished from questions and
commands. The noise of the machine itself can also create difficulties in voice
detection and recognition when the device is in operation during the
conversation.

\subsection{Dialog Strategy}
To achieve natural-seeming interactions, we let users interrupt the agent, and
also let the agent interrupt users. The agent might interrupt a user
mid-question, for example, to announce that a copy job has finished or that a
problem has been detected (e.g. out of paper). In order to support delivery of
unprompted information, the system triggers production rules in response to the
current context and the current utterance. Initiation of a new task, which can
happen even while another one is in progress, pushes the new task onto a stack.

There are several strategies within the dialog manager designed to maximize the
chance of successfully accomplishing tasks [\citet{Kamm10031}]. To address the
problem of discoverability (see Section \ref{sec-challenges}), and to assist
users dealing with unexpected issues [\citet{Myers:2018:PUO:3173574.3173580}],
the agent has the ability to help in understanding the machine and how to
interact with the conversational interface itself. Users can ask for
descriptions of printer concepts and functions including options for each. A
help system is available during the course of a task, and a how-to system can
guide people through an unfamiliar task. The system recognizes when users
encounter obstacles and provides a walkthrough skill where the agent can work
through a process one step at a time. The agent may also employ fallback
strategies that provide more context or information to guide users back on
track. It also has the ability to direct users physical interactions with the
machine. It can answer questions like, ``Where do I find the feeder?'' and
``Where can I find my copied document?''

\subsubsection{Procedures}
The solution understands how to perform procedures.  Accomplishing tasks in
collaboration with a user requires maintaining a dialog state that represents
the knowledge relevant to the current conversation. The state includes a list
of all turns that have occurred up to the current point.  Each turn includes
the utterance and the list of dialog acts that have been parsed from it. Given
a new utterance and its dialog acts, the agent updates its state with the new
information. Only when utterances are unclear or ambiguous, will the agent ask
for confirmation before moving on. Once the state updates are complete, the
agent processes the top task from its stack of goals. Given the current goal,
the agent decides the next thing to ask or say. The agent will ask for any
missing information if it's needed to proceed.  Once all of the necessary
information has been collected for a particular task, it will get a final
confirmation that its understanding of the full job is correct.  Once
confirmed, it initiates the action on the printer and then reports the status
results provided by the printer.

\subsubsection{Diagnosis}
Diagnosis is necessary when users encounter problems during the course of a
procedure. The dialog agent consults a diagnostic engine to get a list of
recommendations to solve a problem. The diagnostic engine might offer a
recommended next step or suggest a condition that should be checked. The
conversational agent determines which recommendation to follow based on the
current state of the dialog. For example, if a solution has already been tried,
it can be eliminated from consideration.  The agent continues to iterate until
all the recommendations have been tried, the problem is fixed, or the user
decides to stop the process.

\section{User Study}
\label{sec-user-study}

We conducted a user study to assess the success of the solution for our target
user population. We also wanted to know how users felt about the solution and
how effective they thought it would be in their work.  Nine of the eleven
individuals from our focus group participated in the usability study. The
specific goals of the study were to
\begin{enumerate}
\item{Assess participants' ability to successfully complete a series of tasks
on an MFP through verbal commands only,}
\item{Quantify quality of experience for ease and a perceived sense of
naturalness of the interaction, and}
\item{Analyze conversational task completion against that of conventional
tap/touch MFP}
\end{enumerate}

PARC's Institutional Review Board approved the protocol for the study.
Participants were allowed to drop out at any time for any reason and indicated
their verbal consent after being read the description of the study and what was
expected of them. They were given gift cards as a thank you for participating in
the focus group and user study.


\begin{figure*}[ht]
\centering
\includegraphics[scale=0.75]{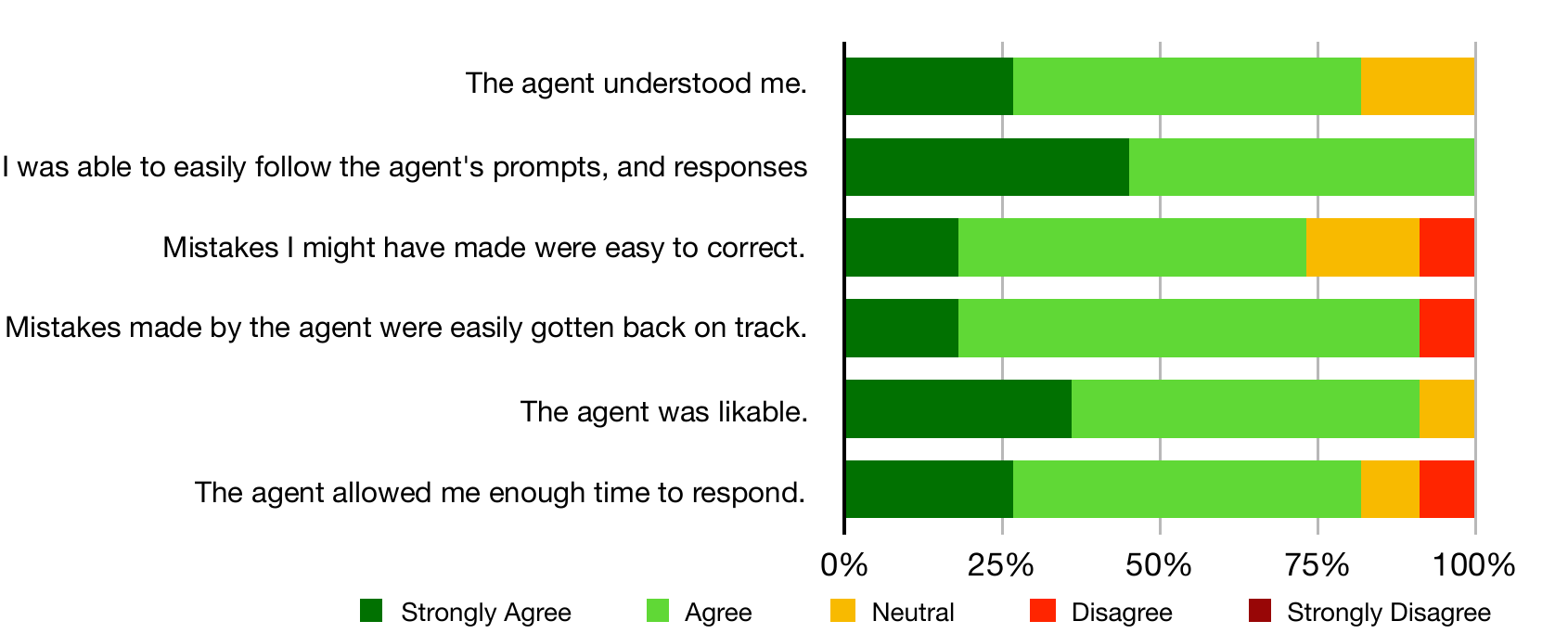}

\caption{Distribution of scores for the first set of survey
questions}
\label{fig-barchart}
\end{figure*}

We did not have a baseline for conventional tap/touch interfaces for
comparison, so data regarding item (3) was collected as part of a pre-test
survey, which was administered before the usability study. The survey collected
data regarding frequency of copier use by task, friction points, workarounds,
task confidence, and task avoidance.

For the study, each participant performed tasks related to seven different
scenarios. A test administrator read aloud each scenario. Once the scenario was
understood, participants used an MFP equipped with our conversational agent to
accomplish the goals suggested by the scenario. They did not have to operate
any other controls on the printer, accomplishing the task through conversation
only. So as not to prime users with the words known to be understood by the
system, scenario descriptions were intentionally written with a task's
obvious ``keywords'' omitted motivating participants to articulate utterances in
their own words.

As an example, the following scenario is designed to prompt a participant to
ask the MFP for three copies of a given document without priming the words
`copy', `document', etc.:
\begin{quote}
\textit{You are organizing a meeting for three coworkers. You want to hand out
the meeting agenda at the meeting. Use the copier to accomplish your goal.}
\end{quote}

At the end of the scenario tests, participants answered questions from a
post-test survey, which collected data regarding their perceptions of
ease-of-use, ability to mitigate error, clarity, precision, benefits of voice
over tactile interaction, frustrations, successes, efficiency, and quality of
the overall experience. The survey included both open-ended questions for
qualitative results and statements where they ranked their level of agreement
for quantitative results.  Two examples of open-ended questions were ``How
would you describe the agent's personality?'' and ``What did you like least
about using the conversational copier?''

For the quantitative statements, participants selected values from a Likert
scale of 1 to 5 (corresponding to Strongly Agree, Agree, Neutral, Disagree and
Strongly Disagree, respectively) for the following statements:

\begin{itemize}
\item{The agent understood me.}
\item{I was able to easily follow the agent's prompts and responses.}
\item{Mistakes I might have made were easy to correct.}
\item{Mistakes made by the agent were easily gotten back on track.}
\item{The agent spoke too fast.}
\item{The agent spoke too slow.}
\item{The agent was able to summarize and confirm my request.}
\item{The agent was repetitive.}
\item{The agent was too wordy.}
\item{The language the agent used was precise.}
\item{The agent was likable.}
\item{I knew how to ask for help if I needed it.}
\item{If I needed help, the agent was helpful.}
\item{At times, I was frustrated or impatient with the agent.}
\item{I knew what to say to initiate a task.}
\item{I knew when a task was successfully completed.}
\item{I knew when the conversation for a task was over.}
\item{The agent allowed me enough time to respond.}
\end{itemize}

The overall success was apparent in the notable contrast between pre- and
post-test survey results. In particular copier-related tasks identified in the
pre-test survey as difficult, time-consuming, or avoided (multiple page
documents, emailing, scanning, stapling, anything more than a simple one-page
copy) were each successfully completed with the use of the conversational
assistant. We present a brief overview of our findings here.

In the first set of sentences, participants were asked about the agent's
behavior. The overwhelming majority of the participants found the agent likable
(91\%) and felt that it understood them (82\%). A significant majority also
thought that the agent allowed them enough time to respond (82\%), and all the
participants agreed that they were able to easily follow the prompts. Also, for
most of the participants, the mistakes made by them or the agent were easy to
correct, while about 10\% of the participants experienced some difficulties
getting back on track. See Figure \ref{fig-barchart} for the distribution of
Likert responses for these statements.

Participants were asked to rate specific aspects of the agent's language. While
a majority of the participants thought the language used by the agent was
precise, about 40\% of them also felt that the agent was repetitive. However,
they insisted that the repetition or confirmation was a positive attribute
in this case, and agreed that the agent was able to summarize and confirm their
request in all tasks. Most of the participants felt the speed of the delivery
was right, while a small portion (about 10\%) felt the agent spoke too slowly.
Also, only a few participants (about 10\%) felt the agent was too wordy.

Participants were then asked if they knew how to ask for help, and if the agent
was helpful when they needed it. About 45\% of the participants responded that
they did not know how to ask for help, while, only 55\% of the participants
felt the agent was helpful when they needed it. Also, about 60\% of the
participants felt frustrated or impatient with the agent at some point during
the test. About 75\% of the participants agreed they knew what to say
to initiate a task. While only 73\% of the participants knew when the
conversation for a task was over, all participants understood when a task
itself was successfully completed.

\section{Conclusion}

We are encouraged by the results of the study and believe that high quality task-based conversational interfaces are possible and can be a great benefit to blind and other users who might otherwise have limited access to different types of technology. There are several technical items that we were not able to accomplish because of the time we had available for this project as well as the limitations created by our architectural design; however, these are solvable given more time and different kinds of technical access to the MFP devices. The bigger problem we hope to address with future research is to reduce the amount of time and expertise required to create these interactions. Modeling of the equipment, the tasks, the users, and the conversations are time-consuming and require people from a variety of disciplines. We are hoping to continue the cross-disciplinary research needed for rich, truly collaborative interactions between people and technology.

\bibliography{mfd_cui}

\end{document}